# Potential Applications of Electrodynamic Tethers

**Sai Charan Petchetti[1]\***

[1]FIITJEE Junior College, Hyderabad, India

**\* Correspondence:**
saicharan@astronomy.org.in

**Abstract**

In recent decades, alternative propulsion systems have been investigated and have attracted great interest in the space community. Most of these alternative propulsion systems need propellants. One alternative propulsion system that does not require propellants and is not often discussed is electrodynamic tethers (EDTs). This paper speculates the potential applications of EDTs ranging from radiation belt remediation to momentum exchange tether re-boosting.

## 1 Introduction

Electrodynamic tethers (EDTs) are long, thin, conductive wires deployed in space that could be used to generate power and thrust. Currents conducted through the wire combined with the strength of the Earth's magnetic field could be harnessed to produce a force that could be used as propellant-less propulsion. The most important benefit of such a system is that it does not require any fuel. Missions such as the Plasma Motor Generator and Tethered Satellite System-1 have demonstrated that EDTs work in space environments. This paper aims to raise awareness of the benefits that EDTs could offer.

## 2 Discussion

### 2.1 A Brief Overview of Electrodynamic Tethers

The idea of using tethers for different purposes in space environments has been a part of the literature for decades. EDTs, however, are a fairly recent innovation. EDTs, at their core, are essentially electrically conductive wires in space that could be used for various purposes, such as radiation remediation and satellite re-boosting. The first research on EDTs was conducted in the 1970s (Lorenzini and Sanmartín, 2004). The shuttle-born era saw some interest in EDTs as an alternative propulsion system. However, unfortunately, development has since been slow.

EDTs are a new and promising technology with great potential. Although multiple past missions have demonstrated the feasibility of using EDTs, EDTs are still in their infancy compared with other similar established technologies. The use of an EDT could have enormous benefits in reducing risks, cost, and fuel requirements and in increasing the efficiency and performance of space missions.

### 2.2 Implementable Solutions

*Table 1. Solutions That Can Be Implemented Using Electrodynamic Tethers (EDTs)*

| Challenges | Time | Cost | Implementable Solutions |
|---|---|---|---|

| Radiation belt remediation | Short term | Low cost | Charged tethers can be used to scatter radiation belt particles into a pitch angle loss cone, leading to the absorption of the particles in the atmosphere. |
|---|---|---|---|
| Satellite De-orbit | Short term | Low cost | Single to multiple EDT systems can be used to de-orbit satellites by generating a Lorentz force drag. |
| Power Generation | Short to medium term | Low cost | EDT systems can be used to convert kinetic energy into electric energy. |
| Satellite Re-boost | Short to medium term | Low to medium cost | Short to medium EDTs can use solar power to push against a planetary magnetic field to achieve propellant-less propulsion. |
| Satellite Attitude Stabilization and Control | Medium term | Medium cost | Multi-electrodynamic tether systems can be used for attitude stabilization, precise pointing, and orbit maneuvering. |
| Momentum Exchange Tether Re-boost | Long Term | High cost | Momentum Exchange Tether Re-boosting using EDTs works similarly to Satellite Re-boosting using EDTs. The key difference is the difference in scale and complexity. |

The Earth's radiation belts can cause mission-threatening radiation damage to both satellites and humans. Therefore, it is necessary to protect passengers and payloads from harmful exposure to radiation particles. Prior studies have often examined methods for mitigating radiation exposure using magnetic shielding. However, magnetic shielding requires large masses and is effective only at certain frequencies. Charged EDTs could be an alternative to magnetic shielding. The tethers work by scattering radiation belt particles into a pitch angle loss cone, leading to the absorption of radiation particles by the atmosphere (Hudoba de Badyn et al., 2016).

EDTs can be used to deliberately decay an object's orbit. Simulations have shown that EDT propulsion technology can be used in near-polar orbits to de-orbit satellites efficiently (Zhu and Zhong, 2011). As part of the TSS-1R mission, a tether system was also used to deliberately gradually decay the orbit of a small satellite to demonstrate electric power generation. Bare EDTs have also been shown to be more mass efficient than their most direct competitor, the Ion Thruster, for re-boosting and de-orbiting objects in orbit (Sanmartín et al., 2008). An EDT designed to de-orbit a 1000-2000 kg spacecraft will likely be about 5-10 km long and would have a mass of 15-30 kg (Sobel and Barcelo, 2007).

To generate electrical energy in space environments, EDTs can be used to convert the kinetic energy of satellites into electrical energy. In the absence of solar power, the combined use of EDTs and chemical rockets is more mass efficient than the direct use of fuel cells (Sanmartín et al., 2008). However, it is important to note that solar power is typically more efficient in such circumstances.

One of the other important challenges that EDTs can help alleviate is re-boosting satellites in space. Artificial satellites that have low-altitude orbits constantly interact with Earth's atmosphere. Although the density of air at such high altitudes is very low, a sufficiently long interaction with air particles could significantly modify the orbit of the satellite, leading to orbital decay. Current solutions used to counter this atmospheric drag range from solid rocket propulsion systems to electric propulsion systems. Almost all these propulsion systems require propellants. Propellants are difficult and expensive to transport and cannot be refueled in most cases. EDT systems can use solar power to push against a planetary magnetic field to achieve propellant-less propulsion. EDT propulsion generates thrust by using a planet as a reaction mass rather than an expelled propellant (Bilén et al., 2012).

EDTs have also been proposed to have the potential to help maintain the International Space Station's orbit. This would work similarly to satellite re-boosting using EDTs. The "International Space Station Electrodynamic Tether Reboost Study" (Johnson and Herrmann, 1998) concluded that the payoff from the use of an EDT in the International Space Station is "considerably greater." The same study estimated that, with a low development and operation cost of only USD 50 million, a tether re-boost system on the International Space Station could potentially save the program up to USD 2 billion over a span of 10 years. An EDT of roughly 20 kilometres in length would be required to power a manned space station. Such a tether is expected to deliver up to 40 kW of electricity, which is adequate for most space stations. (Sobel and Barcelo, 2007).

Stabilization and control of attitude is an essential part of satellite operations. Simulations have demonstrated that multi-electrodynamic tether systems can contribute to attitude stabilization, precise pointing, and orbit maneuvering (Weis and Peck, 2012). This is especially useful for chip-sized spacecraft. A multi-electrodynamic tether system in a chip-sized spacecraft can stabilize the attitude while simultaneously performing orbital maneuvers. Some amount of force and torque control can be exercised in chip-sized spacecraft by directing geomagnetically induced currents. This may help in the passive attitude control of chip-sized spacecraft apart from active attitude control.

Momentum exchange tethers are space tethers that could be used as a launch system and to change spacecraft orbits. EDTs can be integrated into a Momentum Exchange Tether to create a Momentum-exchange/electrodynamic reboost (MXER) facility. MXER facilities have been proposed to boost spacecrafts from a low Earth orbit to a higher orbit like an "upper stage in space" (Canfield et al., 2006). Such a system would rebuild its orbital momentum using EDT thrust; thus, minimizing use of propellants.

## 2.3 Limitations

Despite all their benefits, EDTs have two critical limitations: they are easily severed by the impact of micro-debris and are subject to power surges and vibrations in the tether. It is virtually impossible using current technology to repair an EDT once the tether is severed. The best currently available mitigation strategy is to have redundant current paths to make up in the event of a string failure and to refine electronics and other components of the EDT to better withstand power surges and electrostatic discharges (Sobel and Barcelo, 2007).

Two other aspects that largely determine the complexity of using EDTs are ensuring contact of the tether with the ionosphere and determining the inclination of the orbit.

Regardless of the specifics of an EDT system, the tether must maintain contact with the ambient ionosphere in order to generate propulsion and/or power. One way to achieve the required contact is by using plasma contactors at each end of the EDT. Early experimental efforts in the 1980s have indicated that hollow cathodes and hollow cathode-based plasma sources are sufficient for EDT operations (Patterson and Wilbur, 1986). However, later research has identified that plasma contactors are a bottleneck for EDTs to reach higher performances (Rivo, 2009). Another recently proposed method for electron collection is to leave parts of the tether bare. The inherent advantage of a bare electrodynamic tether is the absence of mass and complexity of contactors (Hastings and Roy, 1993)

EDT operations are also heavily dependent on the inclination of the orbit as the Lorentz force is the vector product of the earth's magnetic induction vector and the current vector. The voltage that is generated by an EDT is dependent on the orbital inclination. For electrodynamic thrust, it is important that the tether is oriented along the radial vector in its orbit (Sanchez and Lozano, 2015) which could limit the manoeuvrability of an EDT.

## 2.4   Summary

EDTs have been demonstrated to have varying levels of success. The differentiating factor between EDTs and most other propulsion technologies is that the former does not require propellant. EDT systems offer great potential by reducing the mass and power requirements for a spacecraft and its maneuvers. However, their use requires further testing and research. This paper draws attention to EDTs and provides a brief review of their potential applications.

## 3   Conflict of Interest

The authors declare that the research was conducted in the absence of any commercial or financial relationships that could be construed as a potential conflict of interest.

## 4   References

[1] Lorenzini, E., and Sanmartín J. (2004) "Electrodynamic Tethers in Space." Scientific American 291, no. 2: 50-57. http://www.jstor.org/stable/26060646.

[2] Hudoba de Badyn, M., Marchand, R., and Sydora, R. D. (2016). Using orbital tethers to remediate geomagnetic radiation belts, J. Geophys. Res. Space Physics, 121, 1114–1123, doi:10.1002/2015JA021715.

[3] Zhu, Z. H. & Zhong, R. (2011). Deorbiting Dynamics of Electrodynamic Tether. International Journal of Aerospace and Lightweight Structures (IJALS). doi:10.3850/2010428611000043.

[4] Sanmartin J., Lorenzini, E., and Manuel Martinez-Sanchez. A Review of Electrodynamic Tethers for Space Applications. AIAA 2008-4595. 44th AIAA/ASME/SAE/ASEE Joint Propulsion Conference & Exhibit. July 2008. doi:10.2514/6.2008-4595.

[5] Sobel, E., & Barcelo, B. (2007). Space Tethers: Applications and Implementations. [Dissertation]. [Worcester, UK] Worcester Polytechnic Institute. https://digital.wpi.edu/concern/student_works/r781wg65d?locale=en.

[6] Bilén, S. G., Stone, N., Scadera, M., Fuhrhop, K. P., Elder, C. H., Hoyt, R. P., Gilchrist, B. E., Alexander, L., Wiegmann, B. M., & Johnson, C. L. (2012). The Propel Electrodynamic Tether Mission and Connecting to the Ionosphere. 12[th] Spacecraft Charging and Technology Conference. https://ntrs.nasa.gov/citations/20120015031.

[7] Johnson, L. & Herrmann, M. (1998). International Space Station Electrodynamic Tether Reboost Study. doi:10.1063/1.57614.

[8] Weis, L., and Peck, M. Attitude Control for Chip Satellites using Multiple Electrodynamic Tethers. (2012). AIAA 2012-4871. AIAA/AAS Astrodynamics Specialist Conference. doi:10.2514/6.2012-4871.

[9] Canfield, S. L., Norris, M. A., & Sorensen, K. F. (2006). Design Rules and Analysis of a Capture Mechanism for Rendezvous between a Space Tether and Payload. NTRS [Preprint]. https://ntrs.nasa.gov/citations/20070001536.

[10] Patterson, M.J., & Wilbur, P.J. (1986). Plasma contactors for electrodynamic tether. NTRS [Conference Paper]. https://ntrs.nasa.gov/citations/19870008995.

[11] Rivo, M. S. (2009). Self Balanced Bare Electrodynamic Tethers. Space Debris Mitigation and other Applications. [Ph.D. Thesis]. [Madrid, Spain] Universidad Politécnica de Madrid. http://oa.upm.es/1839/2/MANUEL_SANJURJO_RIVO.pdf

[12] Hastings E.E., Roy R.I.S. (1993). A Brief Overview of Electrodynamic Tethers. In: De Witt R.N., Duston D., Hyder A.K. (eds) The Behavior of Systems in Space Environment. NATO ASI Series (Series E: Applied Sciences), vol 245. Dordrecht: Springer. doi:10.1007/978-94-011-2048-7_32.

[13] Sanchez, M. M., & Lozano, P. (Spring 2015). 16.522 Space Propulsion. Electrodynamic Tethers. Massachusetts Institute of Technology: MIT OpenCourseWare. https://ocw.mit.edu/courses/aeronautics-and-astronautics/16-522-space-propulsion-spring-2015/lecture-notes/MIT16_522S15_Lecture25.pdf.